\documentclass[12pt]{article}
\usepackage{epsf}
\usepackage{amsmath}
\usepackage{graphics}
\usepackage{cite}

\renewcommand{\thefootnote}{\#\arabic{footnote}}
\begin{document}

\newcommand{\gtrsim}{ \mathop{}_{\textstyle \sim}^{\textstyle >} }
\newcommand{\lesssim}{ \mathop{}_{\textstyle \sim}^{\textstyle <} }

\newcommand{\rem}[1]{{\bf #1}}

\renewcommand{\thefootnote}{\fnsymbol{footnote}}
\setcounter{footnote}{0}
\begin{titlepage}

\def\thefootnote{\fnsymbol{footnote}}

\hfill April 2009

\vskip .5in

\begin{center}

\vskip .5in
\bigskip
\bigskip
{\Large \bf Fermion Mixings in SU(9) Family Unification} 

\vskip .45in

{\bf Paul H. Frampton$^{(a)}$\footnote{frampton@physics.unc.edu}
and Thomas W. Kephart$^{(b)}$\footnote{tom.kephart@gmail.com}}

\bigskip
\bigskip

{$^{(a)}$ Department of Physics and Astronomy, UNC-Chapel Hill, NC 27599.}

{$^{(b)}$ Department of Physics and Astronomy, Vanderbilt University,
Nashville, TN 37235.}

\end{center}

\vskip .4in

\begin{abstract}
In an SU(9) model of gauged family
unification, we propose 
an explanation
for why angles observed in the  
lepton flavor ({\it PMNS}) mixing
matrix are significantly larger
than those measured
for any analagous quark flavor ({\it KM}) mixing angle.
It is directly related to a see-saw mechanism that
we assume to be responsible for the generation
of neutrino masses. Our model is more constrained and 
therefore even more predictive than a model previously proposed by Barr.
\end{abstract}

\end{titlepage}

\renewcommand{\thepage}{\arabic{page}}
\setcounter{page}{1}
\renewcommand{\thefootnote}{\#\arabic{footnote}}

\newpage

\noindent {\it Introduction.}

\bigskip

\noindent When grand unified theories (GUTs)
of strong and electroweak
interactions were at their most popular (from the 1970s
through the 1980s) non-zero neutrino masses had been predicted
but not experimentally established. Thus any effort to address
the neutrino mixing issue was not then motivated for want of  
empirical data.

\bigskip

\noindent Nevertheless, by resurrecting ideas about family
unification which were suggested at that time, in the present
note we shall show that the unexpected nature of neutrino
mixing, especially that two of the neutrino mixing
angles are substantially larger than any of the quark
mixing angles, can be explained.

\bigskip

\noindent The idea of family unification
is to embed a one-family GUT
based on gauge group $G^{'}$ in
a three-family GUT based on a bigger gauge
group $G$, usually with $G^{'} \subset G$.
Sequential family replcation in $G^{'}$
arises from the complex representation
$\rho$ of $G$ and its decompostion $\rho \rightarrow
\rho^{ ~'}$ under $G \rightarrow G^{'}$.

\bigskip

\noindent {\it Minimal SU(9) model of family unification}

\bigskip

\noindent By far the  simplest family unification model
\footnote{A more general family unification scheme is in \cite{Nandi}.
Other $SU(N)$ models are in
\cite{N1,N2,N3,N4,N5,N6,N7,N8,N9,N10,N11,N12}}
is the $SU(9)$ model suggested in 1980 by one of the present
authors\cite{SU9}.

\bigskip

Since the low energy fermions are chiral, the representation
$\rho$ of $G$ must be arranged to be anomaly free. This
is straightforward especially if one restricts attention
to the totally antisymmetric 
irreps $[{\bf k}]$ of rank k for $SU(N)$
as is necessary to avoid color irreps bigger than triplet
and antitriplet. Normalizing the anomaly
$A(N, k)$ such that $A(N, 1)=1$ the values
of $A(N, k)$ satisfy a generalized Pascal triangle
with rows labeled by $N$ (the first row is $N=3$) and columns by $k$
values $1 \le k \le N$):

\begin{center}

1 ~~ -1

1 ~~~ 0 ~~~ -1

1 ~~~ 1 ~~~ -1 ~~~ -1

1 ~~~ 2 ~~~ 0 ~~~ -2 ~~~ -1

1 ~~~ 3 ~~~ 2 ~~~ -2 ~~~ -3 ~~~ -1

1 ~~~ 4 ~~~ 5 ~~~ 0 ~~~ -5 ~~~ -4 ~~~ -1

1 ~~~ 5 ~~~ 9 ~~~ 5 ~~~ -5 ~~~ -9 ~~~ -5 ~~~ -1

\end{center}

\bigskip

\noindent and so on {\it ad infinitum}.

\newpage

\bigskip

\noindent Family unification for three families requires
that when the irreps $(N, k)$ contained in $\rho$ are
decomposed with respect to the standard model
gauge group $G_{SM} \equiv SU(3) \times SU(2) \times U(1))$ 
then $\rho \rightarrow \rho^{'}$ under $G \rightarrow G_{SM}$
leads to a $\rho^{'}$ with three chiral families up
to possible $G_{SM}$-singlets and {\it real}
representations of $G_{SM}$. These real representations
can all pair up into Dirac mass terms and generally
acquire masses inaccessible to present colliders.
We shall refer to all such states as {\it superheavy}.

\bigskip

\noindent The appropriate decomposition is facilitated by 
employment of $SU(5)$ which is equivalent for the
group theory to the standard model
\footnote{This does not imply that $SU(5)$ is a good symmetry
at any energy.}
and the decomposition then involves binomial coefficients
$C^M_q \equiv N! [q! (N-q)!]^{-1}$ according to
\begin{equation}
\rho \equiv (N, k) \rightarrow \rho^{'} \equiv \Sigma_p C^{N-5}_p (5, k-p)
\label{decomposition}
\end{equation}

\bigskip
\bigskip
\bigskip

\noindent There is one more issue, cancellation
of anomalies in $\rho$, which automatically implies anomaly-freedom
and complete families in the standard model.
Writing $A(N, k)$ to represent the chiral anomaly of representation
$(N, k)$ normalization such that the defining representation
$(N, 1)$ has anomaly $A(N, k) = +1$ one employs the well-known
result
\begin{equation}
A(N, k) = \left[ \frac{(N-3)! (N-2k)}{(N-k-1)!(k-1)!} \right]
\label{anoamly}
\end{equation}
which yields the generalized Pascal triangle mentioned above. Anomaly freedom
for $\rho \equiv \Sigma_p B_p (N, p)$ is assured if and only if
$\Sigma B_p A(N, p) = 0$.

\bigskip
\bigskip

\noindent The result of searching the possible $\rho$
for each $N$ leads to the conclusion reached three decades ago
that the simplest family unification occurs for $N=9$ and
the model is \cite{SU9}
\begin{equation}
(9, 3) + 9 (9,1)^{*}
\label{SU9one}
\end{equation}
which may be rewritten with $N=9$ suppressed as
\footnote{Recall that $[{\bf k}]$ denotes a totally
antisymmetric $k^{th}$-rank irrep of $SU(9)$.}
\begin{equation}
[{\bf 3}] + 9 [{\bf 1}]^{*}
\label{SU9two}
\end{equation}
with dimensions ${\bf 84} + 9 ({\bf \bar{9}})'s$. 
One may perhaps best mnemonicize it by 
\begin{equation}
{\bf 9}^3 + 9 ({\bf \bar{9}}).
\label{SU9three}
\end{equation}

\bigskip

\noindent When this $SU(9)$-model was built
\footnote{The smallest appropriate {\it irreducible}
representation of $SU(N)$ occurs for $N=6$
with dimension \cite{EKK,OP,PS} greater than $3 \times 10^5$
so the reducibility in Eqs.(\ref{SU9one},\ref{SU9two},\ref{SU9three})
is inevitable.}
in 1979, there was little experimental
evidence for neutrino mass and absolutely nothing was known about PMNS
mixing. At present comparably as much is known about PMNS as about
the much longer-studied KM quark mixing matrix.

\bigskip

\noindent In what follows we shall use the $SU(9)$-model to 
attempt to explain   
the significant difference between 
observed lepton and quark mixings.

\bigskip
\bigskip
\bigskip

\noindent {\it Group Theoretic Representations}

\bigskip
\bigskip

\noindent Let us examine more carefully 
the fermion fields extant in the $SU(9)$-model.
The group-theoretic bookkeeping is facilitated by the use of a $SU(5)$
subgroup of $SU(9)$, yet at any point we may rewrite in the
standard model $SU(3)_C \times SU(2)_L \times U(1)_Y$
subgroup and there is no implication whatsoever that
$SU(5)$ is a symmetry or that the additional twelve
cofactor gauge bosons exist physically. I.e, we could reduce the 
$SU(9)$-model to a $SU(3)_C \times SU(2)_L \times U(1)_Y\times SU(4)_family$
model. However, keeping the full $SU(9)$ imposes constraints 
among coupling constants and other parameters.

\bigskip

\noindent We introduce the notation $A, B, C, .. = 1-9$ for $SU(9)$;
$I, J, K,..= 1-5$ or $SU(5)$; and $k, l, m,..=1-4$ for the $SU(4) \subset SU(9)$
which commutes with the $SU(5) \subset SU(9)$.

\bigskip
\bigskip

\noindent The $(9,3)$ fermions in Eq.(\ref{SU9one}) can thereby be rewritten
\begin{equation}
\Psi^{ABC} \equiv \Psi^{IJK} + \Psi^{IKk} + \Psi^{Ikl} + \Psi^{klm}
\label{84}
\end{equation}
while the $9(9, 1)^{*}$ fermions of Eq.(\ref{SU9one}) become
\begin{equation}
9 \Psi_A \equiv 9 \left( \Psi_I + \Psi_k \right)
\label{81}
\end{equation}

\bigskip

\noindent The quarks and leptons of the three-family standard model
and their respective mixings are not difficult to read off from Eqs.
(\ref{84}) and (\ref{81}).

\bigskip

\noindent To keep track of quarks and leptons,
we first recall the locations of the flavor eigenstates
in the first fermion family. The second and third familes
are {\it mutatis mutandis}.

\bigskip

\noindent We denote the three QCD colors as Red (R), Green (G) and Blue (B).

\bigskip

\noindent Of these fifteen chiral fermions, ten are located in 
\begin{equation}
\Psi^{IJ} \equiv \left( 
\begin{tabular}{ccc|cc}
0 & $\bar{u}^B$ & - $\bar{u}^G$ & - $u^R$ & -$d^R$ \\
- $\bar{u}^B$ & 0 & $\bar{u}^R$ & - $u^G$ & -$d^G$ \\
$\bar{u}^G$ & - $\bar{u}^R$ & 0 & - $u^B$ & -$d^B$ \\
\hline
$u^R$ & $u^G$ & $u^B$ & 0 & - $e^{+}$ \\
$d^R$ & $d^G$ & $d^B$ & $e^{+}$ & 0
\end{tabular}
\right)
\label{10}
\end{equation}

\bigskip

\noindent and the remaining five are in

\begin{equation}
\Psi_I \equiv \left( \bar{d}^R, \bar{d}^G, \bar{d}^B | \nu_e, e^{-} \right)
\label{5}
\end{equation}

\bigskip

\noindent In Eqs. (\ref{10}) and (\ref{5}) the $SU(3)_C$ and $SU(2)_L$
factors of the standard model gauge group are indicated. 

\bigskip

\noindent Weak hypercharge $Y$ defined by $Q = (T_{L3} + \frac{1}{2} Y)$ corresponds 
to the historically normalized $SU(5)$ generator

\begin{equation}
Y = {\rm diag} \left( - \frac{2}{3}, - \frac{2}{3}, - \frac{2}{3} | +1, +1
\right)
\label{Y}
\end{equation}

\bigskip
\bigskip
\bigskip
\bigskip

\noindent {\it Quark Mixings}

\bigskip
\bigskip

\noindent Keeping renormalizability, hence avoiding
irrelevant operators, quark mixings will arise from
two Yukawa couplings

\begin{equation}
\lambda  \Psi^{ABC}\Psi^{DEF}H^{GHI} \epsilon_{ABCDEFGHI}
\label{84s}
\end{equation}

\bigskip

\noindent and

\bigskip

\begin{equation}
\lambda^{'}_{(b)} \Psi^{ABC}\Psi_A^{(b)}H_{BC}  
\label{36}
\end{equation}

\bigskip

\noindent where indices in parentheses run over the nine $ {\bf \bar{9}}$s.
At the $ SU(5)\times SU(4)$ level, these terms reduce to 

\bigskip

\begin{equation}
\lambda \Psi^{IJa}\Psi^{KLb}H^{Mcd} \epsilon_{IJKLM}\epsilon_{abcd}
\label{10105}
\end{equation}

\bigskip

\noindent and

\bigskip

\begin{equation}
\lambda^{'}_{(\hat b)} \Psi^{IJa}\Psi_I^{(\hat b)}H_{Ja}  
\label{1055}
\end{equation}

\bigskip

\noindent where $\hat b = 1, 2, 3$ is a family label for those $ {\bf \bar{5}}$s of $SU(5)$
that stay light.

\bigskip

\noindent When the Higgs doublet acquires a VEV in the
neutral component $H_4$, there result mass matrices $U$
and $D$ for up- and down- quarks respectively.
One way to proceed is to identify the
flavor and mass eigenstates for the up-quarks
then diagonalize
\begin{equation}
{\cal D} \equiv D D^{\dagger}
\label{calD}
\end{equation}
by
\begin{equation}
V_{KM}^{\dagger}{\cal D}V_{KM} = {\rm diag}(M_b^2, M_s^2, M_d^2)
\label{KM}
\end{equation}

\bigskip

\noindent To zeroth order in the Cabibbo angle
$\beta = \sin \Theta_{12} \sim 0.22$, the $KM$ matrix
$V_{KM}$ is a unit matrix. The three mixing angles
in $V_{KM}$ are all small, not more than 0.22 radians.

\bigskip

\noindent Thus, the Yukawa couplings in Eq.(\ref{10105})
and Eq.(\ref{1055}) are approximately diagonal in family
space. Although it could have been otherwise, Nature
chooses Yukawa couplings with this property.

\bigskip

\noindent The big question is: once we accept that neutrinos
are massive and there exists a lepton counterpart $V_{PMNS}$
to the quark $V_{KM}$ mixing, why is it not similar to $V_{KM}$?

\bigskip

\noindent When the neutrino angles $\theta_{ij}$ were
measured it seemed a surprise to many in the theory community
that the atmospheric  and solar angles, $\theta_{23}$ and
$\theta_{12}$, are larger than $\Theta_{12}$. In
retrospect, could this have been a anticipated?

\bigskip
\bigskip
\bigskip
\bigskip

\noindent {\it Lepton Mixings}

\bigskip
\bigskip

\noindent The Yukawa coupling in Eq.(\ref{1055}) gives
a mass matrix for the charged leptons
\begin{equation}
L_{(\hat b)}  = \lambda^{'}_{(\hat b)}<H_4>
\label{L}
\end{equation}

\bigskip

\noindent For the neutrinos we adopt the idea that they
have Majorana (not Dirac) masses. Then there must be 
right-handed neutrino fields $N_R^n$ where the
label $n$ is usually taken to be $n=1, 2, 3$
one for each family. It is also possible
that there are only two right-handed neutrinos
$n=1,2$ that contribute to the sea saw with the consequence that one mass eigenvalue
vanish.

\bigskip

\noindent The neutrino mass matrix $M_{\nu}$ is
assumed to arise\cite{Minkowski} from a see-saw
mechanism
\begin{equation}
M_{\nu} = M_D M_N^T M_D^T
\label{seesaw}
\end{equation}

\bigskip

\noindent The underlying couplings are
\begin{equation}
\left[ (M_D)_{an} \nu^a N_R^n + h.c. \right] + (M_N)_{nn'} N_R^n N_R^{n'}
\label{nucouplings}
\end{equation}

\bigskip

\noindent It is general to adopt a basis where the charged
leptons have degenerate flavor and mass eigenstates
whereupon the lepton mixings $V_{PMNS}$ are contained in
\begin{equation}
V_{PMNS}^{\dagger}{\cal N}V_{PMNS} = {\rm diag}(m_3^2, m_2^2, m_1^2)
\label{PMNS}
\end{equation}

\bigskip

\noindent where we have introduced the matrix ${\cal N}$ by

\noindent

\begin{equation}
{\cal N} \equiv M_{\nu} M_{\nu}^{\dagger}
\label{NU}
\end{equation}

\bigskip

\noindent Of course, we know the answer for $V_{PMNS}$
but let us objectively scrutinize the see-saw in Eq.(\ref{seesaw})
with the repeated index summation convention

\bigskip

\begin{equation}
(M_{\nu})_{ab} = (M_D)_{an} (M_N^T)_{nn'} (M_D^T)_{n'b}
\label{indices}
\end{equation}

\bigskip

\noindent and for general $(M_N^T)_{nn'}$ which
involves arbitrary mixing between the $N_R^n$, 
there is no reason for $(M_{\nu})_{ab}$ to be approximately
diagonal. Consequently there is every reason for
the lepton mixings in $V_{PMNS}$ to be of order
one, not small like the quark
mixings in $V_{KM}$.

\bigskip

\noindent In the $SU(9)$-model, according to Eqs.(\ref{84},\ref{81})
there are as many as fourty chiral fermions without
3-2-1 charge. The right-handed neutrinos $N_R^n$ with
$n = 1,2$ are among these. All are superheavy
but only two participate in the see-saw mechanism
with the three light left-handed neutrinos.

\bigskip

\noindent We note that without breaking $SU(4)$
eight states can be paired up according to

\begin{equation}
\Psi^{klm}\Psi_n\epsilon^n_{klm}
\label{epsilon}
\end{equation}

\bigskip

\noindent Once $SU(4)$ is broken to $SU(2)$ all of the
remaining thirty-two singlet chiral fermions
can successfully acquire superheavy masses.

\bigskip

\newpage

\bigskip
\bigskip

\noindent {\it Larger Mixing Angles for Leptons than for Quarks}

\bigskip

\noindent In $SU(9)$ the qualitative difference between neutrino and quark
mixings arises from the different $SU(4)$ dependences
appearing in Eqs. (\ref{10105}) and (\ref{1055}), for quarks, which
are all $SU(4)$ non-singlets and
in Eq. (\ref{epsilon}), for right-handed neutrinos, which are all $SU(4)$ singlets.

\bigskip

\noindent As already pointed out after Eq. (\ref{KM}), the matrix $V_{KM}$
has small off-diagonal elements corresponding to the fact
that Nature chooses simple Yukawa couplings for quarks 
which are almost flavor diagonal.
For leptons, on the other hand,
the see-saw mechanism Eq. (\ref{seesaw})
leads to the neutrino mass matrix Eq. (\ref{indices}) which
is not simply related to 
Yukawa couplings, but instead is more closely related to the Dirac and Majorana
mass terms in Eqs. (\ref{nucouplings}) and (\ref{NU})
respectively. This underlies why the lepton mixing angles
are larger than the quark angles using
the difference between $SU(4)$ transformation
properties evident in our Eqs. (\ref{84}) and (\ref{81})
from reference \cite{SU9}.

\bigskip
\bigskip

\noindent {\it Discussion}

\bigskip
\bigskip

\noindent We have seen that one could have anticipated
before the measurements that some neutrino mixing angles
would be significantly larger than any quark
mixing.

\bigskip

\noindent It could be of some interest to investigate
fermion mixings in extensions
of the standard model such as 
the chiral color model\cite{axigluon} or 
in the
presence of a fourth family\cite{FHS}.

\bigskip

\noindent An $SU(8)$ model has been proposed by Barr\cite{Barr}
with the similar aim. Our model improves on it 
because we use a simpler family unification.

\bigskip

\noindent Furthermore it is not necessary to limit 
ourselves to non-supersymmetric models.
In \cite{Frampton:1982mj} it was shown that 
models exist with supersymmetric family unification. 
$SU(8)$ and $SU(9)$ models were explored and 
gauge symmetry breaking was carried out that 
preserved supersymmetry \cite{Frampton:1981pf} 
to a low scale. While the particle content is 
somewhat different in these models, the basic 
conclusion stays the same. We expect a low energy 
theory, here the MSSM, extended and constrained 
by a gauged family symmetry broken at an 
intermediate scale.

\bigskip

\noindent As our final open question we ask: 
Can a discrete flavor symmetry be successfully 
embedded in a gauged family unification model? For example,
the $SU(4)$ gauged symmetry is contained within $SU(9)$
and  commutes with the standard model
gauge group. This $SU(4)$ contains $SU(2)$ subgroups
which have the binary tetrahedral
group as a subgroup. Can such
a binary tetrahedral subgroup in the present 
model be identified with the 
flavor symmetry used in {\it e.g.} \cite{FK,FK2,FKM}?
Numerous other models with continuous non-abelian family 
symmetries exist \cite{collection}, and the same question 
can be posed for all such models.

\vspace{2.0in}

\begin{center}

{\bf Acknowledgements}

\end{center}

\bigskip

\noindent This work was supported by U.S. Department 
of Energy grants number DE-FG02-06ER41418 
and  DE-FG05-85ER40226.

\end{document}